# How is science discussed on Bluesky?

Er-Te Zheng[1], Xiaorui Jiang[1], Zhichao Fang[2,3*], Mike Thelwall[1*]

* Corresponding authors

Er-Te Zheng (ORCID: 0000-0001-8759-3643)
[1] School of Information, Journalism and Communication, The University of Sheffield, Sheffield, UK.
E-mail: ezheng1@sheffield.ac.uk

Xiaorui Jiang (ORCID: 0000-0003-4255-5445)
[1] School of Information, Journalism and Communication, The University of Sheffield, Sheffield, UK.
E-mail: xiaorui.jiang@sheffield.ac.uk

Zhichao Fang (ORCID: 0000-0002-3802-2227)
[2] School of Information Resource Management, Renmin University of China, Beijing, China.
[3] Centre for Science and Technology Studies (CWTS), Leiden University, Leiden, The Netherlands.
E-mail: z.fang@cwts.leidenuniv.nl

Mike Thelwall (ORCID: 0000-0001-6065-205X)
[1] School of Information, Journalism and Communication, The University of Sheffield, Sheffield, UK.
E-mail: m.a.thelwall@sheffield.ac.uk

## Abstract

Amid the migration of academics from X, the social media platform Bluesky has emerged as a potential alternative. To assess its viability and relevance for science communication, this study presents the first large-scale analysis of scholarly article dissemination on Bluesky, exploring its potential as a new source of social media metrics. We collected and analysed over 2.6 million Bluesky posts referencing 532,302 scholarly articles from January 2023 to July 2025, integrating metadata from the OpenAlex database. Temporal trends, disciplinary coverage, language use, textual characteristics, and user engagement were examined. A sharp increase in scholarly activity on Bluesky was observed from November 2024 to January 2025, coinciding with broader academic shifts away from X. As on X, Bluesky posts primarily concern the health, social, and environmental sciences and are predominantly written in English.

Nevertheless, Bluesky posts demonstrate substantially higher levels of interaction (likes, reposts, replies, and quotes) and greater textual originality than previously reported for X, suggesting both stronger interactive and more interpretive engagement. These findings highlight Bluesky's emerging role as a credible platform for science communication and a promising source for altmetrics. The platform may facilitate not only early visibility of research outputs but also more meaningful scholarly dialogue in the evolving social media landscape.

## 1. Introduction

Over the past two decades, science communication has expanded beyond the traditional confines of academic publishing and entered the dynamic and participatory spheres of digital media (Brossard & Scheufele, 2013; Peters et al., 2014). The rise of social media platforms has played a transformative role in this shift, enabling the rapid and widespread dissemination of scientific knowledge to broader audiences, while simultaneously fostering public engagement with scientific discourse (Coletti et al., 2022; Ortega, 2018; Yu et al., 2019; L. Zhang et al., 2023). This evolution has not only democratised access to scholarly information to a much greater degree than was previously achieved by science journalism, but has also prompted the development of new metrics, practices, and evaluative frameworks for assessing the visibility of scholarly outputs. Among these developments, altmetrics – and more specifically, social media metrics – have emerged as complementary indicators of attention for scholarly outputs on digital platforms (Costas, 2017; Haustein et al., 2016; Priem et al., 2012; Sugimoto et al., 2017).

A persistent challenge in leveraging social media metrics for research evaluation lies in the fluid and volatile nature of the platforms themselves (Haustein, 2016). The social media ecosystem has undergone considerable shifts, as exemplified by the decline or discontinuation of once-prominent services such as MySpace, Vine, and Google+, and the concurrent emergence of decentralised or niche platforms – such as Mastodon and Bluesky – that offer alternative spaces for scholarly exchange and public dialogue (Cava et al., 2023; McIntyre, 2014). As the digital communication landscape continues to evolve, it is necessary to investigate how these emerging platforms contribute to the circulation, reception, and engagement with scientific knowledge. Understanding these dynamics is crucial not only for tracking the diffusion of research but also for adapting science communication strategies and evaluative practices in a rapidly changing media environment.

### *1.1 Social media studies of science communication*

A wide array of platforms – including Twitter (now X), Facebook, Reddit, and Mendeley – have been examined as venues for monitoring and understanding science communication in online environments (Thelwall, Haustein, et al., 2013). These platforms also serve as key data sources for prominent altmetric aggregators such as Altmetric, PlumX, and Crossref Event Data (Erdt et al., 2016; Robinson-García et al., 2014; Zahedi & Costas, 2018). Despite differences in functionality, community norms, and user demographics, social media activities capture forms of scientific discourse that extend beyond conventional academic outlets, such as journals and conferences. As such, they provide valuable supplementary evidence for evaluating the visibility, engagement, and potential societal impact of scholarly outputs (Brossard & Scheufele, 2013).

Among these platforms, Twitter/X has been the most extensively studied, due in large part to the volume, accessibility, and structured nature of its science-related data. Estimates suggest that Twitter/X once accounted for over 80% of all social media mentions of scholarly publications (Peng et al., 2022). Furthermore, more than one-third of Web of Science-indexed publications previously received at least one mention on Twitter/X (Fang et al., 2020), but this figure may have shrunk alongside increasing concerns about the platform and its new owner.

Research on Twitter/X mentions of scholarly works has explored a wide range of dimensions, including temporal trends, geographic distribution, disciplinary variation, and correlations with traditional citation impact. In terms of timing, Twitter/X attention tends to spike immediately following an article's publication, reflecting the platform's rapid information turnover and real-time communication norms (Ortega, 2018; Shuai et al., 2012). Geographically, the bulk of Twitter/X activity originates from users based in the United States, Europe, and Japan (Yu et al., 2019). Disciplinary analyses have revealed that research in health sciences, social sciences, and environmental sciences often receives disproportionately high levels of attention, possibly due to the public relevance and accessibility of such topics (Fang et al., 2020; Haustein et al., 2015).

In addition, a substantial body of literature has examined the relationship between Twitter/X activity and citation performance. While most studies report a positive but weak correlation between the two (Bornmann, 2015; Costas et al., 2015; Haustein et al., 2014; Zahedi et al., 2014), some have identified moderate to strong correlations, which may be attributed to differences in sample size, field-specific practices, or methodological design (Eysenbach, 2011; Hayon et al., 2019; Peoples et al., 2016; Vaghjiani et al., 2024).

### *1.2 Bluesky as an emerging platform for science communication*

The acquisition of Twitter by Elon Musk and its subsequent rebranding as X introduced a series of transformative changes to the platform, including the implementation of paid features, diminished content moderation, and an increase in the visibility of misinformation (Caulfield, 2023; McKee et al., 2024; Sethi et al., 2025). These developments have led to considerable dissatisfaction among users and prompted a migration of scientists to alternative platforms (Bisbee & Munger, 2025; Vidal Valero, 2023), including Bluesky (Kupferschmidt, 2024; Mallapaty, 2024).

Bluesky was initially launched in 2019 as a research initiative within Twitter but became independent in 2021 (Campbell, 2021). While retaining a user interface and interaction model similar to X, Bluesky distinguishes itself through its adoption of the AT Protocol – a decentralised, open architecture that enables greater transparency, user control, and accessibility for data collection and export (Brown et al., 2024). In recognition of its growing relevance, Bluesky has recently been integrated into Altmetric’s attention tracking system (Kidambi, 2024b).

Alongside growing interest, research into Bluesky’s role in science communication has begun to emerge. For instance, Sethi et al. (2025) provided practical guidance for nephrologists transitioning to Bluesky, while Quelle et al. (2025) analysed 300,000 academic users from Twitter/X and found that 18% had already migrated to Bluesky. Meanwhile, Arroyo-Machado et al. (2025) conducted a comparative study of scholarly engagement on Bluesky and X, focusing on articles published in seven multidisciplinary and Library and Information Science journals between January 2024 and March 2025. While they observed a rise in Bluesky activity – especially after November 2024 – the study’s limited journal coverage constrains the generalisability of its findings. Therefore, there remains a critical lack of large-scale, systematic investigations into how scholarly articles are disseminated on Bluesky, and what characterises this form of interaction.

### *1.3 Research questions*

This study addresses the lack of large-scale analyses of Bluesky by investigating the temporal dynamics, disciplinary distribution, textual features, and user engagement associated with Bluesky posts referencing scholarly articles. By exploring the flow of scientific information on this emerging decentralised platform, the study aims to contribute to a broader understanding of the evolving ecology of social media-based science communication in the post-Twitter era. To guide this investigation, the study poses the following research questions (RQs), each of which is relevant to the potential use of Bluesky posts for altmetrics, as further discussed in the paper:

- **RQ1**. What are the temporal trends in Bluesky posts referencing scholarly articles? For example, are these posts predominantly concentrated around the publication date of the article?
- **RQ2**. Which disciplines and topics are most frequently represented in Bluesky posts referencing scholarly articles?
- **RQ3**. What are the key textual characteristics – such as language use and textual similarity to the original article – of Bluesky posts referencing scholarly articles?
- **RQ4**. To what extent do Bluesky users engage with scholarly posts through behaviours such as liking, reposting, replying, and quoting?

## 2. Data and methods

### *2.1 Data collection of Bluesky posts referencing scholarly articles*

To investigate the dissemination of scholarly articles on Bluesky, we implemented a systematic three-stage data collection procedure, as illustrated in Figure 1. First, we retrieved Bluesky scholarly post URLs from the Altmetric Explorer, covering posts published between 00:00:00 on 1 January 2023 and 23:59:59 on 31 July 2025, along with the DOIs of the scholarly articles mentioned in these posts. This query yielded an initial dataset of 2,696,806 Bluesky scholarly post URLs, referencing 547,514 unique article DOIs. Altmetric was selected as our data source because it was the first platform to systematically track research-related conversations on Bluesky, having added

Bluesky to its list of monitored social media sources in December 2024 (Kidambi, 2024a).

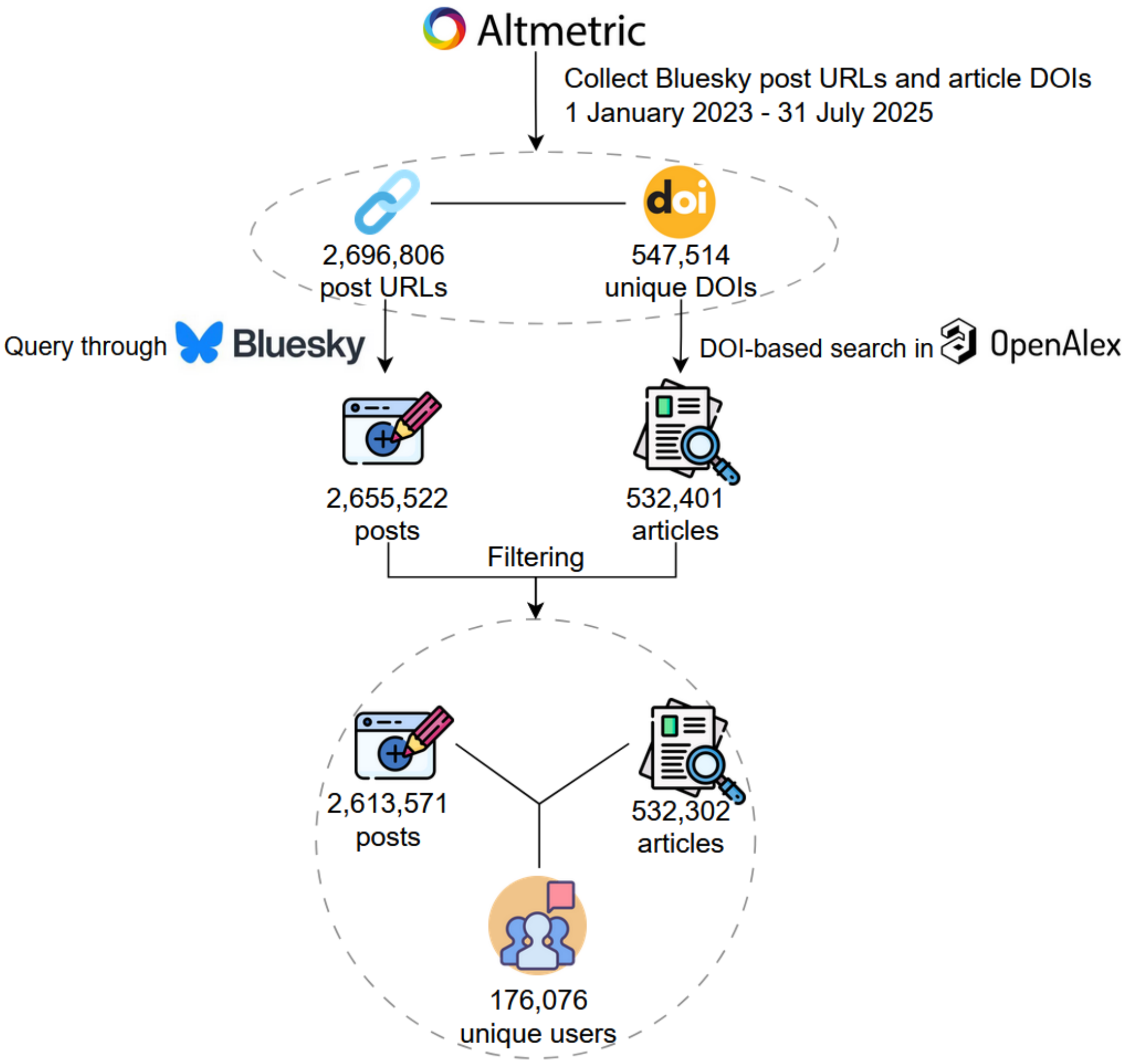

**Figure 1**. Workflow of the Bluesky data collection process.

Second, we used the collected Bluesky post URLs to retrieve post metadata via the official Bluesky API. This process yielded a refined dataset of 2,655,522 Bluesky scholarly posts. The absence of some records is likely due to users deleting their posts after they were initially captured by Altmetric. For each post, we retrieved the timestamp, full textual content, and user engagement metrics. In parallel, we retrieved article-level metadata from OpenAlex, a large-scale, open-access bibliographic resource. OpenAlex offers coverage of scholarly publications with DOIs that is comparable to, or even broader than, that of Web of Science and Scopus (Alperin et al., 2024; Culbert et al., 2025; Haupka et al., 2024), making it a suitable choice for comprehensive bibliometric analysis. On 2 August 2025, all extracted DOIs were queried through the OpenAlex API, which verified 532,401 DOIs as valid scholarly articles. For each matched article, we retrieved the title, publication date, and disciplinary classification at the topic, subfield, field, and domain levels, according to OpenAlex's hierarchical taxonomy.

Following metadata validation, we filtered the original Bluesky posts to retain only those referencing DOIs successfully matched in OpenAlex, along with the corresponding referenced articles. This filtering step produced a final dataset comprising 2,613,571 posts authored by 176,076 unique users and citing 532,302 articles, resulting in 2,675,933 post-article pairs, as a single post may reference multiple articles. For each post, we compiled the following information:

- Post ID: A unique identifier for the Bluesky post.
- User ID: A unique identifier for the user who authored the post.
- Post time: The timestamp of the post's creation.
- Post content: The full text of the post.
- User engagement metrics: The number of likes, reposts, replies, and quotes the post received.

### *2.2 Disciplinary classification of scholarly articles*

To analyse the disciplinary distribution of scholarly articles referenced on Bluesky, we adopted the article-level classification system provided by OpenAlex.[1] OpenAlex constructs a citation network by linking indexed scholarly works through direct citation relationships and algorithmically clusters these works into over 4,000 research areas – referred to as *topics* – using the Leiden algorithm (Traag et al., 2019). These topics are then organised into a four-tiered hierarchical structure comprising 252 *subfields*, 26 *fields*, and 4 broad *domains*, allowing each article to be assigned a classification at the domain, field, subfield, and topic levels.

For analytical clarity and tractability, this study primarily utilises domain-level and field-level classifications to examine overall disciplinary trends, while topic-level classifications are used to identify specific areas of research that receive the most attention on Bluesky. The distribution of scholarly articles across domains and fields is summarised in Table 1, based on the OpenAlex taxonomy.

[1] See more information of the article-level classification system of OpenAlex at: https://help.openalex.org/hc/en-us/articles/24736129405719-Topics (accessed 8 May 2025).

**Table 1**. Distribution of scholarly articles referenced on Bluesky across domains and fields from January 2023 to July 2025.

| Domain | Field | Number of articles | Proportion |
|---|---|---|---|
| Physical Sciences (138,698, 26.4%) | Environmental Science | 41,496 | 7.9% |
| | Engineering | 21,353 | 4.1% |
| | Computer Science | 17,839 | 3.4% |
| | Physics and Astronomy | 17,539 | 3.3% |
| | Earth and Planetary Sciences | 13,459 | 2.6% |
| | Materials Science | 9,396 | 1.8% |
| | Chemistry | 9,318 | 1.8% |
| | Mathematics | 4,313 | 0.8% |
| | Energy | 3,133 | 0.6% |
| | Chemical Engineering | 852 | 0.2% |
| Health Sciences (131,341, 25.0%) | Medicine | 115,512 | 22.0% |
| | Health Professions | 12,361 | 2.4% |
| | Nursing | 2,073 | 0.4% |
| | Veterinary | 765 | 0.1% |
| | Dentistry | 630 | 0.1% |
| Social Sciences (131,098, 25.0%) | Social Sciences – General | 68,829 | 13.1% |
| | Psychology | 25,112 | 4.8% |
| | Arts and Humanities | 14,615 | 2.8% |
| | Economics, Econometrics and Finance | 11,951 | 2.3% |
| | Business, Management and Accounting | 5,836 | 1.1% |
| | Decision Sciences | 4,755 | 0.9% |
| Life Sciences (123,566, 23.6%) | Biochemistry, Genetics and Molecular Biology | 63,680 | 12.1% |
| | Agricultural and Biological Sciences | 24,380 | 4.6% |
| | Neuroscience | 23,645 | 4.5% |
| | Immunology and Microbiology | 10,339 | 2.0% |
| | Pharmacology, Toxicology and Pharmaceutics | 1,522 | 0.3% |

*Note*: Disciplinary classifications were unavailable for 7,599 articles (1.4%), primarily due to insufficient citation data, which precluded their assignment to research areas by OpenAlex. These articles were excluded from the disciplinary analysis.

### 2.3 Content analysis of Bluesky posts

To better understand how scholarly articles are discussed on Bluesky, we conducted a content analysis of post texts from two perspectives: language use and content originality.

To examine language use, we employed the *langs* field provided by Bluesky to classify the primary language of each post. The Bluesky Social client automatically detects the languages present in a post and sets them as the default *langs* value, while users can manually adjust this setting on a per-post basis to more accurately reflect the actual language used.[2] If a post was classified by Bluesky as containing multiple languages, we incremented the count of each detected language by one, treating all listed languages equally in the distribution. This enabled a systematic assessment of linguistic diversity and provided insights into potential regional or cultural variations in science communication practices on Bluesky. Posts for which the language could not be reliably inferred were assigned an empty *langs* field by Bluesky. These 335,613 posts (12.8%) were excluded from the analysis, leaving a final dataset of 2,277,958 posts for the language-use analysis.

To assess content originality, we first cleaned each post by removing @usernames, #hashtags, and URLs, producing a refined textual representation. Following this cleaning procedure, 26,165 Bluesky posts (1.0%) were empty and were therefore excluded, yielding a final corpus of 2,587,406 posts for analysis. We then computed cosine similarity scores between each cleaned Bluesky post and the title of its referenced scholarly article. Both texts were first transformed into TF-IDF vector representations using the *TfidfVectorizer* from the *scikit-learn* Python package, which captures the relative importance of terms within the corpus and allows for comparisons beyond exact word matches. The resulting sparse matrices were then used to calculate cosine similarity scores for each post-title pair.

Cosine similarity is a widely used metric in textual analysis that quantifies lexical overlap while normalising for text length. Given that social media posts often echo or paraphrase article titles (Didegah et al., 2018; Na, 2015; Sergiadis, 2018; Thelwall, Tsou, et al., 2013), measuring textual similarity in this way helps distinguish posts that closely mirror the title from those offering more original or interpretive content. The cosine similarity score, which ranges from 0 to 1, serves as a proxy for assessing the degree of novelty or interpretation offered by Bluesky users. Posts with high similarity may simply reproduce article titles, whereas those with lower similarity scores are more likely to reflect user-generated commentary, summarisation, or reinterpretation, thus offering a deeper understanding of the communicative and interpretive roles users play in disseminating scholarly content.

[2] Further information on Bluesky's language identification system is available at: https://docs.bsky.app/blog/create-post (accessed 7 August 2025).

## 3. Results

### *3.1 Temporal characteristics of Bluesky posts*

(1) Temporal trends in scholarly Bluesky activity

Figure 2 presents monthly trends in the number of Bluesky posts referencing scholarly articles, the number of unique users posting such content, and the number of distinct articles mentioned. From January 2023 to September 2024, all three indicators – posts, users, and referenced articles – remained relatively stable at low levels, with monthly counts below 3,000. This early period of limited activity suggests that Bluesky had not yet been widely adopted for science communication, likely reflecting its nascent stage within academic communities. It also predates Altmetric's systematic tracking of Bluesky, meaning that the subsequent sharp increase in recorded activity likely reflects both genuine uptake and the commencement of Altmetric's indexing. During this time, established platforms such as X likely continued to dominate the science communication landscape.

In October 2024, all indicators showed a modest rise: posts reached around 10,000, users 3,200, and referenced articles 6,500. A substantial increase followed from November 2024 onwards, when scholarly posts surged to around 239,000, accompanied by growth in the number of users (35,000) and referenced articles (54,000). This rapid escalation continued through December and peaked in January 2025, with approximately 324,000 posts by 45,000 users referencing 71,000 unique articles. This pattern aligns with observations by Kupferschmidt (2024), who reported that researchers and scientific institutions began migrating en masse to Bluesky during this period, driven by dissatisfaction with X and the growing recognition of Bluesky as a more credible, open venue for scholarly discourse. The influx of academic users appears to have catalysed a sharp rise in science-related activity on the platform.

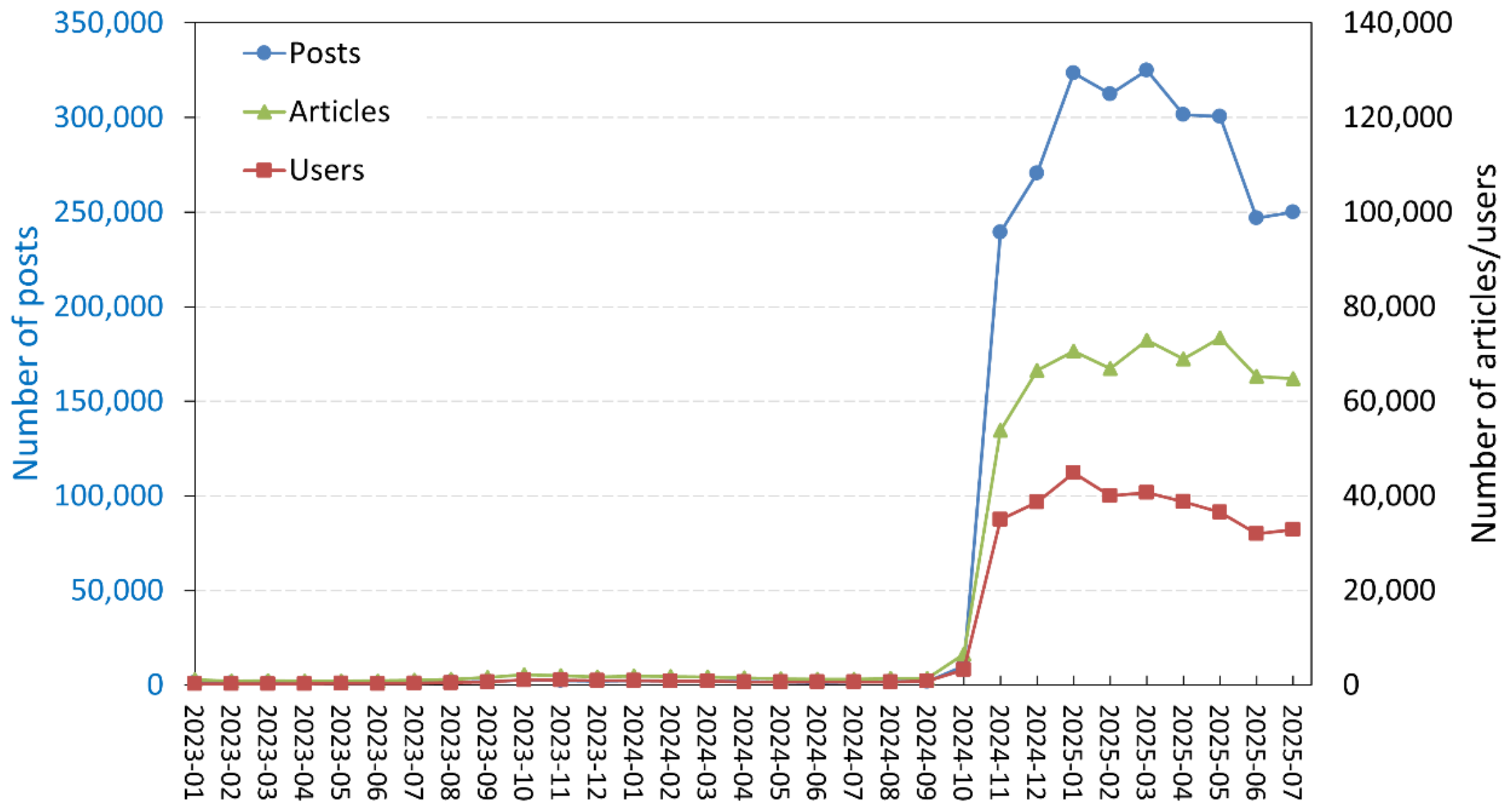

**Figure 2**. Number of Bluesky posts, users, and articles referenced from January 2023 to July 2025.

Altmetric began collecting Bluesky posts on 24 October 2024 (Altmetric team, 2025) and, although it claims to have incorporated historical posts, we sought to determine whether the observed surge in Figure 2 was primarily attributable to this indexing change or to a genuine influx of academic users. To investigate this, we employed an alternative retrieval strategy, using the Bluesky API to collect all posts containing valid article DOIs – indexable in OpenAlex – embedded within their links, as described in the Appendix. The results shown in Figure A1 confirmed that the number of Bluesky scholarly posts rose sharply between November 2024 and January 2025. Similarly, in a case study of seven journals, Arroyo-Machado et al. (2025) used the Bluesky API to gather post data and found a substantial increase in posts referring to these journals' articles during the same period. Taken together, these findings indicate that the main driver of growth in scholarly posts was the large-scale migration of academic users to Bluesky, although the magnitude of the increase was likely amplified by the onset of Altmetric's indexing.

Following the January 2025 peak, scholarly activity between February and July 2025 stabilised at historically high levels before experiencing a slight decline. This modest downturn may reflect the natural subsiding of the initial post-migration surge or a plateau in user acquisition as the platform approached saturation among early academic adopters. Nonetheless, activity remained far higher than pre-surge levels. As noted by Kupferschmidt (2025), Bluesky's reputation for fostering a "good boring" environment – favouring thoughtful, evidence-based discussion over sensationalism – likely contributed to its sustained appeal among researchers seeking focused engagement. This enduring appeal, coupled with the platform's transition from rapid diffusion to steady participation, suggests that Bluesky has consolidated its position as a stable and credible hub for science communication.

(2) Time lag between article publication and post release

To assess the responsiveness of Bluesky users to newly published research, we calculated the *time lag* between an article's publication date (as recorded in OpenAlex) and the date it was shared on Bluesky. Figure 3 shows the distribution of these time intervals, aggregated at the monthly level.

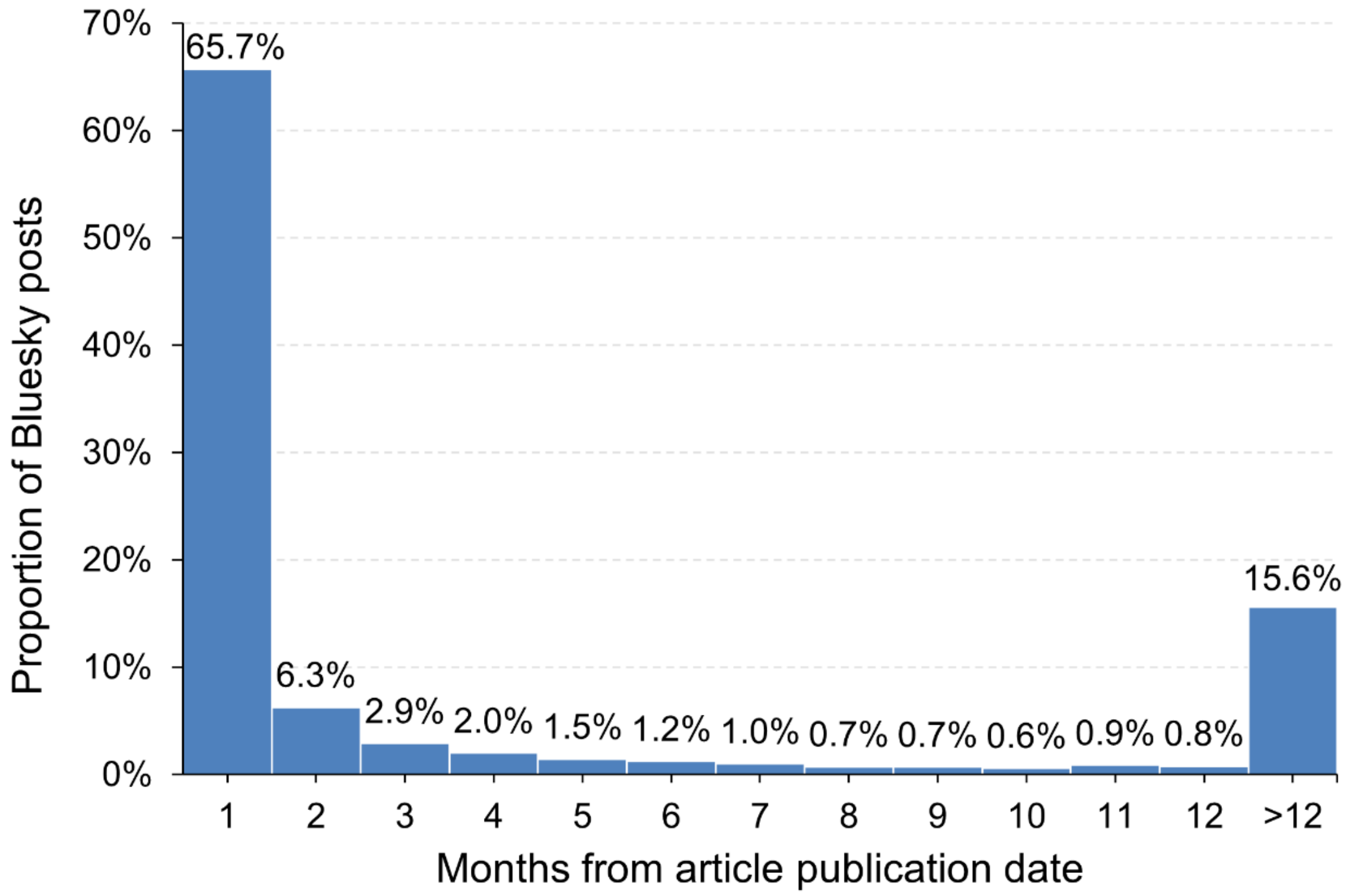

**Figure 3**. Distribution of time lags between article publication and Bluesky post sharing from January 2023 to July 2025. *Note*: To account for potential discrepancies in publication dates, 40,593 post-article mentions (1.5%) that preceded the official publication date of the corresponding articles were classified as occurring within one month after publication.

Approximately 50.7% of posts referencing scholarly articles appeared within the first week of publication, and 65.7% were posted within the first month. These figures highlight Bluesky's potential as a platform for the timely dissemination of scientific knowledge, with nearly two-thirds of scholarly mentions occurring shortly after publication. This responsiveness suggests active engagement by a user base attuned to new research outputs.

At the same time, the data also reveal a considerable degree of delayed engagement: approximately one-third of posts were made after the first month, and 15.6% were posted one year or more after the article's publication. This pattern reflects a balanced temporal dynamic – while immediate attention is common, the platform also supports longer-term interaction with scholarly content. Such a dual pattern suggests that Bluesky functions both as a space for rapid scholarly exchange and as a repository for sustained academic engagement. Alternatively, some new users may have reposted older key papers (e.g., in posts copied from X) when first joined the platform, such as to introduce themselves.

*3.2 Disciplinary variations in Bluesky posts referencing scholarly articles*

The distribution of Bluesky posts referencing scholarly articles across the 26 fields within the four broad OpenAlex domains reveals that, while overall attention levels are relatively consistent across domains, they vary substantially among fields within each domain (Figure 4). The Health Sciences domain accounts for the largest share, comprising 29.0% of all Bluesky posts. Within this domain, Medicine overwhelmingly predominates (25.6%), followed by Health Professions (2.9%). Fields such as Nursing, Veterinary Science, and Dentistry receive relatively limited attention.

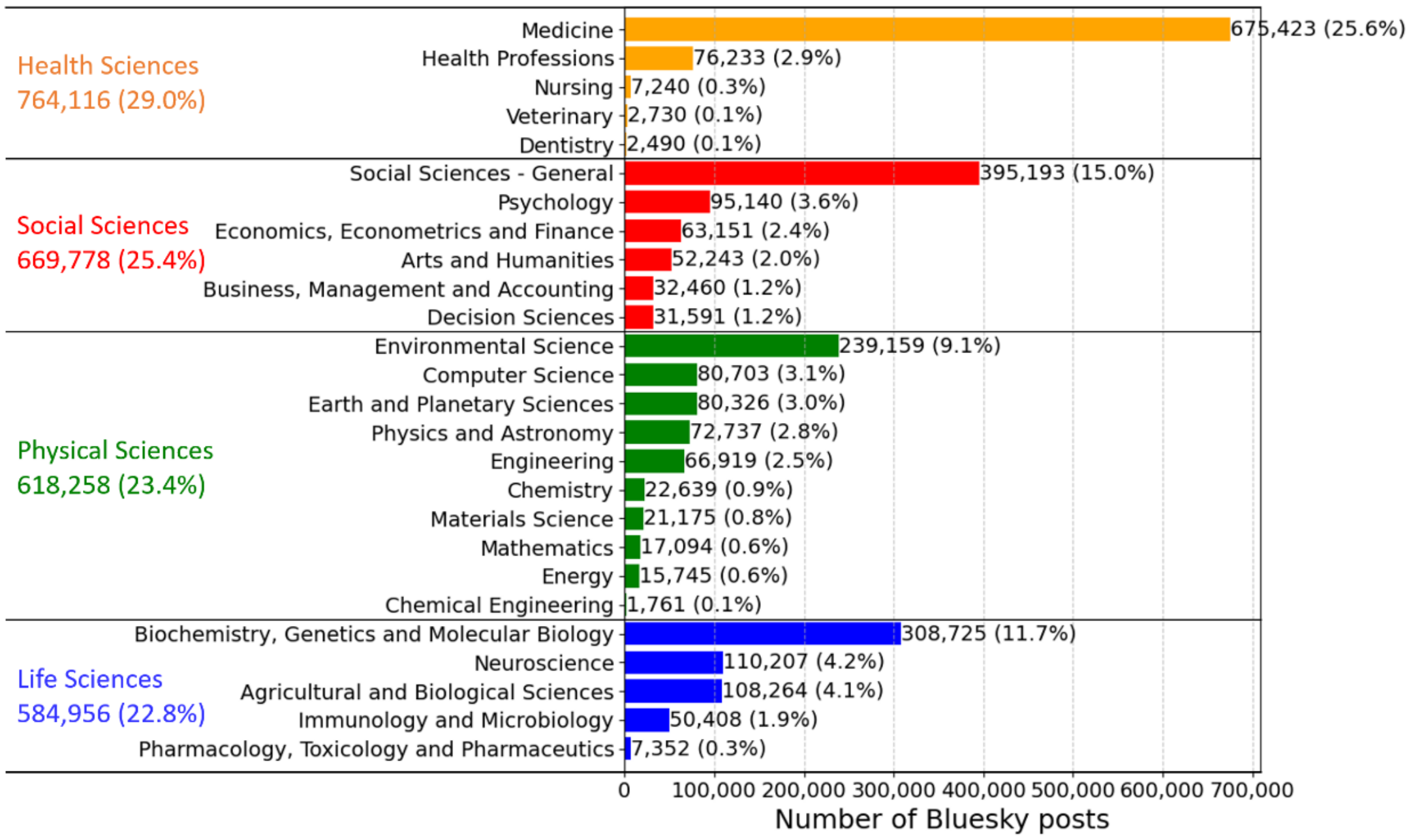

**Figure 4**. Distribution of Bluesky posts referencing scholarly articles across OpenAlex domains and fields from January 2023 to July 2025.

To provide a more fine-grained view of disciplinary engagement, Figure 5 shows the topic-level distribution within each domain, using a scatter plot based on the number of posts and unique users referencing each topic.[3] The broadly linear shapes indicate that a few OpenAlex topics receive disproportionate attention from individual users. The three most prominent topics within each domain – those situated nearest to the upper-right corner of each panel – are annotated to indicate the areas receiving the highest levels of attention. In the Health Sciences, discussions are predominantly oriented towards advances in scientific and medical knowledge, the implications of health and medical research, and the enduring clinical and public health challenges posed by the long-term effects of COVID-19. These thematic emphases indicate that Bluesky users engaging with health-related scholarship are attuned not only to disciplinary developments, but also to their broader impact and persistent global health concerns.

[3] Detailed data on the number of posts and unique users referencing each article topic are openly available at: https://doi.org/10.6084/m9.figshare.29867201.v1.

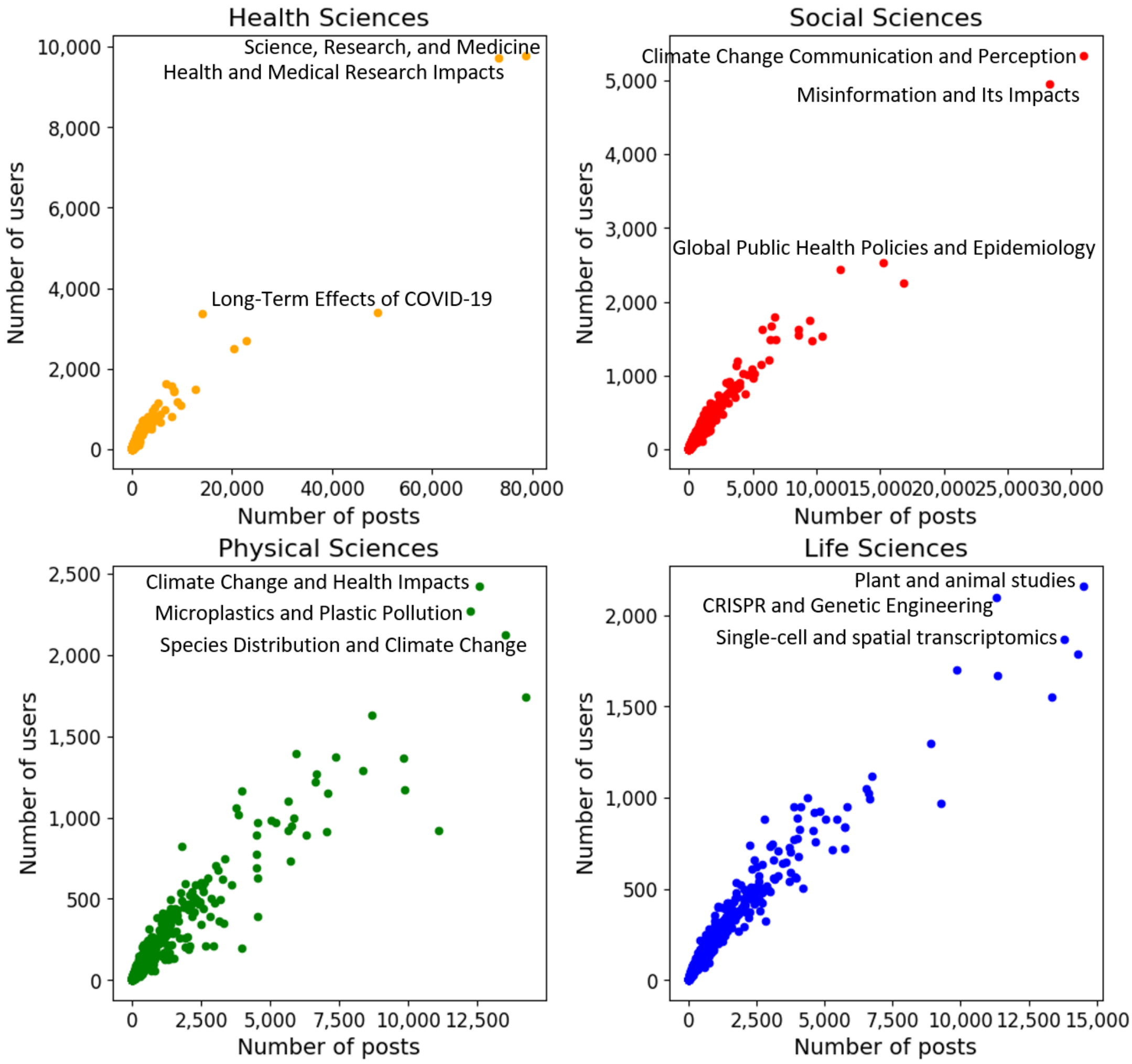

**Figure 5**. Scatter plot of topics of scholarly articles mentioned on Bluesky from January 2023 to July 2025, disaggregated by domain.

The Social Sciences domain accounts for 25.4% of all Bluesky posts. Within this domain, Social Sciences – General is the most referenced field (15.0%), especially in subfields such as Sociology and Political Science, Political Science and International Relations. This is followed by Psychology (3.6%) and Economics, Econometrics and Finance (2.4%). In the Social Sciences, the leading topics tend to address how climate change is communicated and understood by the public, the origins and societal consequences of misinformation, and the formulation and implementation of global public health policies alongside epidemiological research. Taken together, these topics indicate that Bluesky conversations in this domain often gravitate towards complex issues that link environmental sustainability, information integrity, and population health.

The Physical Sciences domain contributes 23.4% of the total posts. The most highly referenced fields are Environmental Science (9.1%), Computer Science (3.1%), and Earth and Planetary Sciences (3.0%). As shown in Figure 5, the strong visibility of Environmental Science reflects global public interest in issues such as climate change-related health impacts, microplastic pollution, and species distribution under changing climatic conditions – topics that resonate with both academic and general audiences.

While the Life Sciences domain represents a smaller proportion of posts (22.8%), it is largely dominated by Biochemistry, Genetics and Molecular Biology (11.7%), followed by Neuroscience (4.2%) and Agricultural and Biological Sciences (4.1%). This distribution indicates ongoing interest in biodiversity and organismal studies, genetic engineering, and the biological mechanisms underpinning life and health, highlighting the broad relevance of these fields. Notably, topics related to plants, animals, and biodiversity – closely tied to environmental sustainability – receive the highest levels of attention. Overall, public health, climate change, and broader environmental concerns represent cross-cutting topics that attract sustained attention from Bluesky users across diverse scientific domains.

### *3.3 Textual characteristics of Bluesky posts referencing scholarly articles*

(1) Linguistic analysis of Bluesky posts

A substantial majority (91.0%) of posts are written in English, reaffirming its role as the *lingua franca* of global science communication (Figure 6). German is the second most frequently used language, accounting for 2.4% of posts, followed by French (2.0%) and Spanish (1.7%). Other languages, including Japanese, Portuguese, Dutch, Italian, and Finnish, each represent less than 1% of the total, indicating limited linguistic diversity in scholarly discourse on Bluesky.

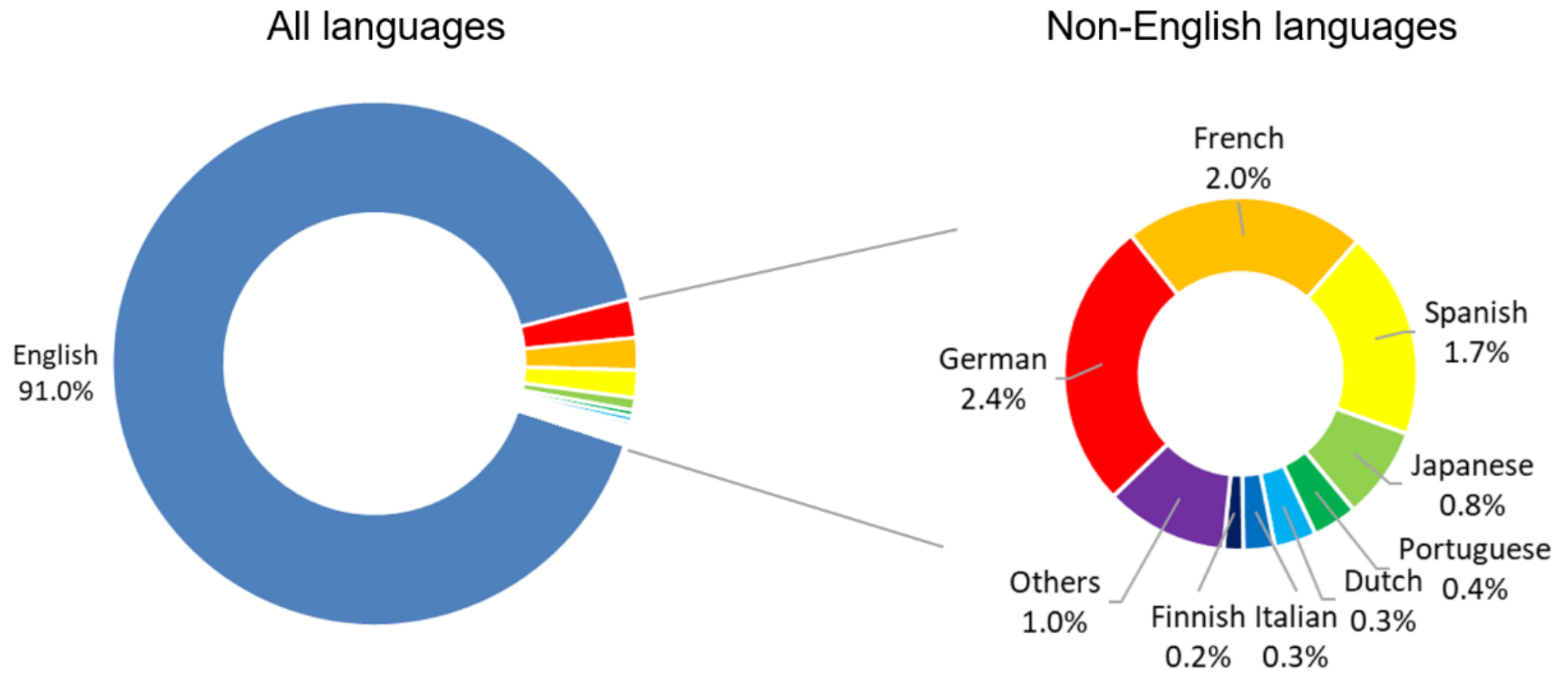

**Figure 6**. The proportion of each language used in Bluesky posts referencing scholarly articles from January 2023 to July 2025.

This linguistic distribution likely reflects both the demographic profile of Bluesky's user base and the entrenched dominance of English in global academic exchange. While English clearly prevails, the presence of German, French, and Spanish posts suggests that some degree of non-English science communication is emerging on the platform. These minority-language contributions underscore Bluesky's growing international reach, while also highlighting the ongoing challenge of fostering truly multilingual engagement in social media-based science communication.

(2) Similarity between post texts and article titles

The majority of Bluesky posts show low textual similarity to the titles of the scholarly articles they reference (Figure 7). Nearly 60% have a cosine similarity score below 0.3, with the remaining distribution tapering off across higher similarity bins. This skewed distribution towards lower similarity values indicates that most posts are not direct reproductions of article titles. Instead, users may tend to introduce, reinterpret, summarise, or comment on the research, reflecting a more active and interpretive form of science communication (as exemplified in Table 2). This behaviour underscores the platform's potential to foster meaningful public discourse around academic research, as users contribute original perspectives or contextual framing rather than passively relaying bibliographic information.

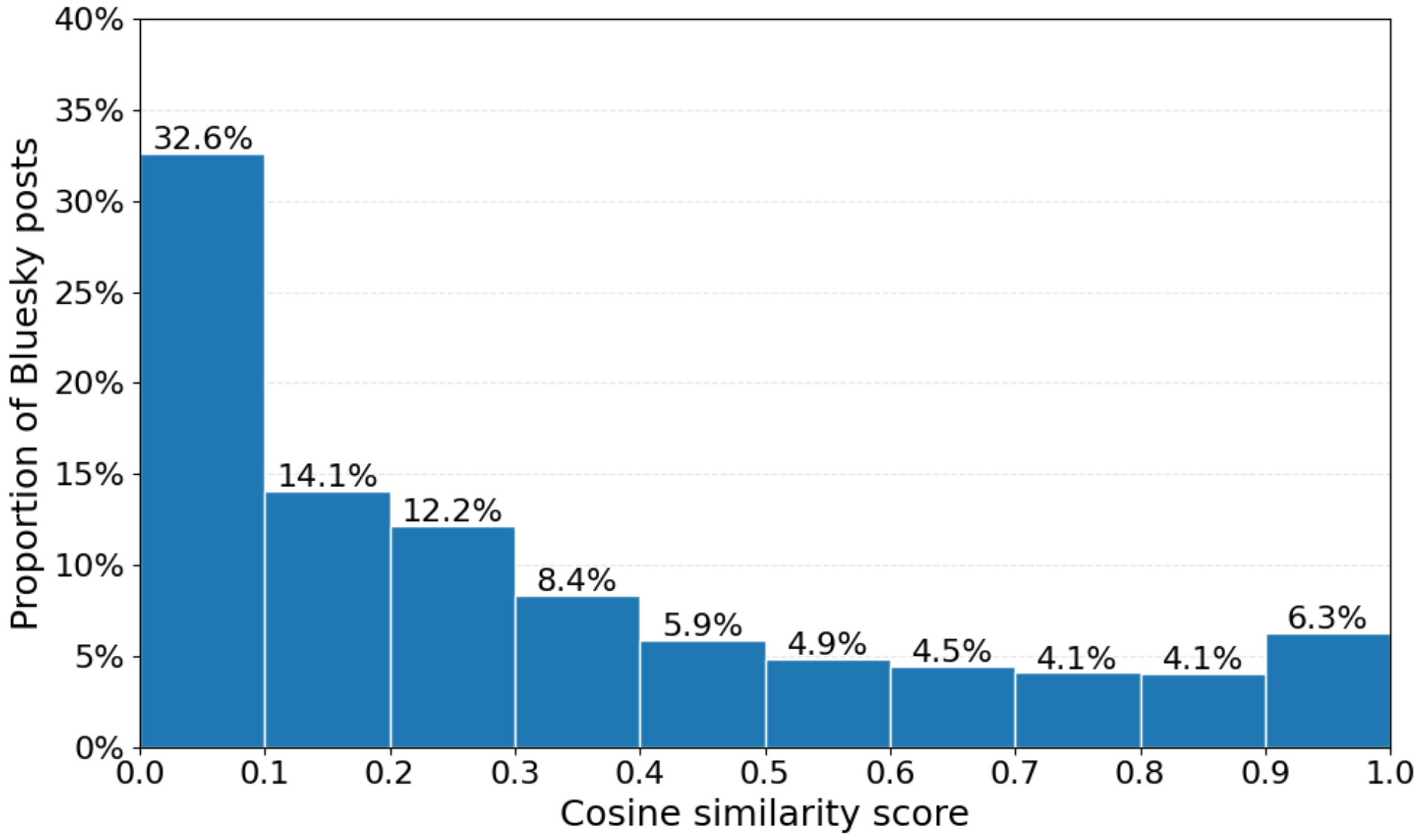

**Figure 7**. Distribution of cosine similarity scores between post text and article titles.

**Table 2**. Sample Bluesky posts with different levels of similarity score to article title.

| **Similarity score** | **Sample Bluesky post** | **Type** |
|---|---|---|
| 0.8-1.0 | The paper *[Title]*, published in *[Journal]* | Title repetition |
| 0.6-0.8 | This study can be a reference for future research *[Title]* | Comment |
| 0.4-0.6 | A proud cross-disciplinary effort between geneticists, zooplankton ecologists & climate scientists! *[Title]* | Introduction |
| 0.2-0.4 | How does increasing international trade truly impact jobs at the local level? In our new open-access publication in *[Journal]*, we tackle this question by providing a detailed analysis of *[Country]*'s local labour markets. | Reinterpretation |
| 0-0.2 | Interesting insights reveal why academics consider leaving the profession. University management should focus on key retention factors: the prospect of a permanent job, overall job satisfaction, and the provision of tenure for women with children. | Summary |

*Note*: Post content has been paraphrased and anonymized to protect user privacy.

### *3.4 User engagement analysis*

We assessed user engagement with science-related posts on Bluesky by analysing the complementary cumulative distribution functions of four types of user engagement metrics: likes, reposts, replies, and quotes, as shown in Figure 8.

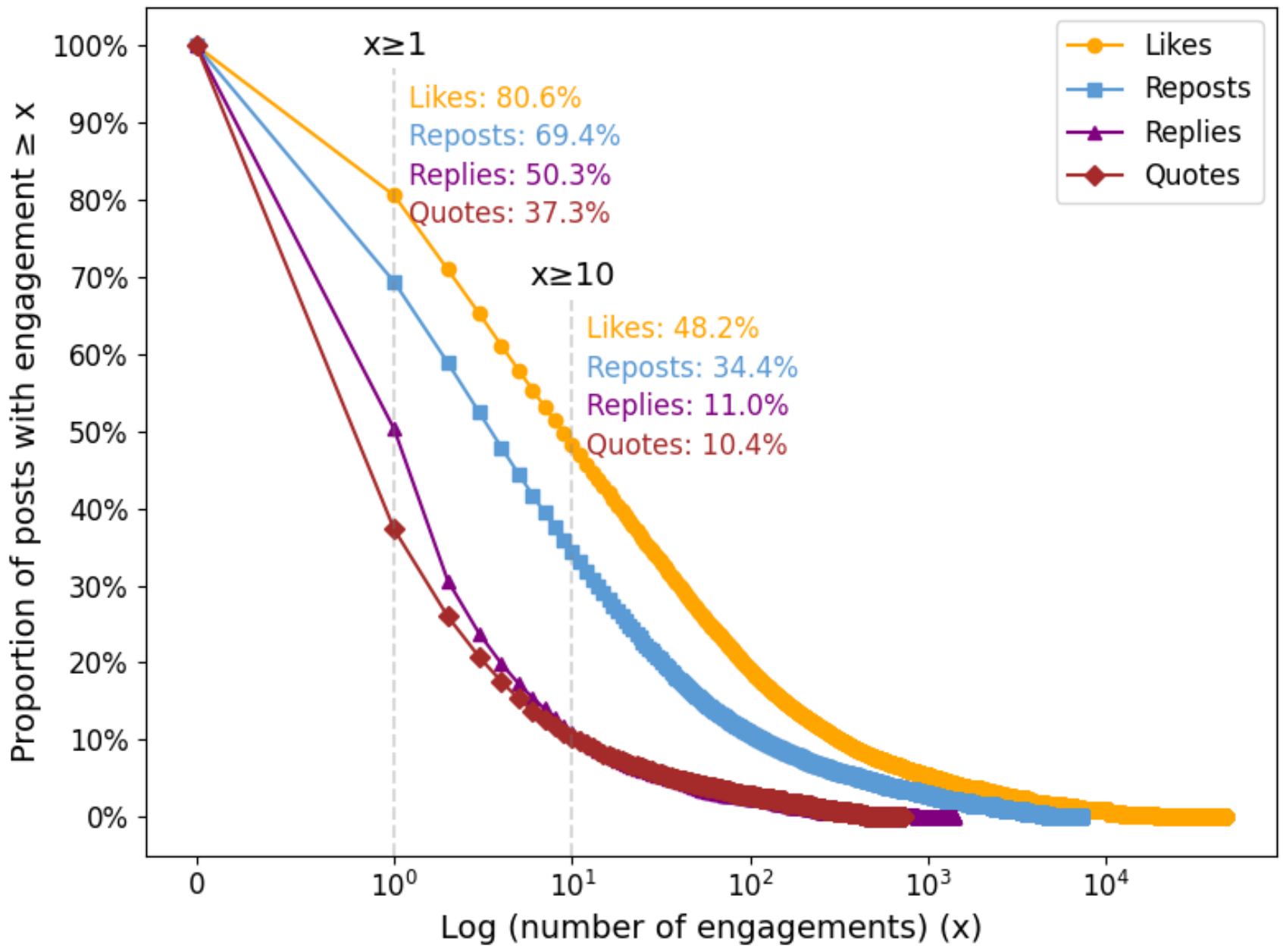

**Figure 8**. Proportion of scholarly Bluesky posts from January 2023 to July 2025 with different levels of engagements.

At the lowest non-zero threshold (interactions ≥ 1), 80.6% of scholarly posts received at least one like, followed by 69.4% with one or more reposts, 50.3% with replies, and 37.3% with quotes. This distribution indicates that a substantial proportion of scholarly content on Bluesky attracts engagement across multiple interaction types. Lower-effort interactions – such as liking and reposting, which require only a single click – are the most common, yet higher-effort forms of participation involving original content creation, such as replies and quotes, also occur at notably robust levels.

When the engagement threshold is set at ten interactions or more, the proportion of posts engaging audiences at this level remains considerable: 48.2% of posts receive at least ten likes, while 34.4% are reposted ten or more times. Replies and quotes at this higher threshold, though less frequent, are still present in 11.0% and 10.4% of posts, respectively, suggesting that in-depth, content-generating interactions are not uncommon within Bluesky's scholarly discourse. Overall, these patterns point to a relatively high baseline of engagement on Bluesky, with a broad share of scholarly posts attracting sustained and diverse forms of audience interaction.

## 4. Discussion

### *4.1 The emergence of Bluesky as an active platform for scholarly discourse*

This study presents an exploratory analysis of Bluesky posts referencing scholarly articles between January 2023 and July 2025, offering early insights into Bluesky's potential as an emerging platform for social media-based science communication and altmetrics.

Bluesky experienced a significant increase in scholarly activity in November 2024, coinciding with political developments in the United States – particularly Elon Musk's involvement in the presidential election – and the resulting exodus of scientists from X (Kupferschmidt, 2024). This migration likely positioned Bluesky as a major destination for academics seeking a more stable, decentralised alternative for scholarly discourse. The trend aligns with findings from Arroyo-Machado et al. (2025), who observed a comparable increase in Bluesky engagement with major multidisciplinary journals and several Library and Information Science journals. Scholarly activity continued to grow until January 2025, after which it stabilised, suggesting that Bluesky has matured into a stable and credible venue for science communication. Its growing relevance was further underscored by its inclusion in the Altmetric tracking system at the end of 2024 (Kidambi, 2024b).

According to Altmetric, Bluesky hosted more posts linked to new research than X on most days in March 2025 (Taylor & Areia, 2025). This highlights Bluesky's substantial potential as a platform for science communication. Functionally, Bluesky closely resembles Twitter/X, and its scholarly use mirrors several well-documented patterns observed on that platform, including:

- Rapid post-publication engagement, with most posts appearing shortly after article publication (Ortega, 2018; Shuai et al., 2012; Yu et al., 2017);
- Disciplinary concentration, with heightened activity around topics in health, social, and environmental sciences (Didegah et al., 2018; Erdt et al., 2016; Haustein et al., 2015);
- Linguistic dominance of English as the primary language of scholarly discourse

(Hassan et al., 2021; Yu et al., 2018; Y. Zhang & Fang, 2025);

- Greater prevalence of low-effort interactions (e.g., likes and reposts) than conversational forms of engagement, such as replies and quotes (Fang et al., 2022).

These parallels demonstrate that Bluesky, like X, is capable of capturing early signals of research visibility. However, it also inherits disciplinary and sociolinguistic biases common in science communication on mainstream social platforms.

Despite these similarities, Bluesky has two notable distinctions in scholarly communication. First, it demonstrates a substantially higher level of interactive engagement than has been reported in prior studies of X. Earlier research on X found that the proportion of tweets receiving at least ten interactions varied by discipline, ranging from 3.9-7.5% for likes, 1.4-4.4% for reposts, 0-0.2% for quotes, and 0-0.1% for replies (Fang et al., 2022). In contrast, our findings show that 48.2% of Bluesky scholarly posts received at least ten likes, 34.4% received ten or more reposts, 10.4% received ten or more quotes, and 11.0% received ten or more replies. The prevalence of likes and reposts on Bluesky is more than an order of magnitude higher than that previously reported for X, while quotes and replies occur at rates approaching two orders of magnitude greater, indicating a markedly higher level of active engagement with scholarly content on the emerging platform.

Second, Bluesky shows a higher level of content originality in scholarly discourse. Our analysis of cosine similarity between post texts and article titles revealed that only 6.3% of Bluesky posts replicate article titles nearly verbatim (i.e., cosine similarity between 0.9 and 1). In contrast, prior studies of X reported title replication rates ranging from 11.8% to 92.4%, varying by discipline (Didegah et al., 2018; Kumar et al., 2019; Na, 2015; Sergiadis, 2018; Thelwall, Tsou, et al., 2013). Bluesky was also found to have a higher proportion of original posts and a lower proportion of reposts compared to X (Taylor & Areia, 2025). This contrast suggests that, whereas X has primarily served as a dissemination tool (Htoo & Na, 2017), Bluesky may support a more interpretive, reflective mode of science communication.

This greater originality may be attributed to Bluesky's evolving user base, which – particularly after the migration from X – may comprise a higher proportion of academic users actively engaged in scholarly discourse. In contrast, X's broader and more heterogeneous audience tends to favour simple retweets or title-sharing behaviours.

These findings position Bluesky not only as a viable successor to X but also as a potentially richer environment for substantive science communication.

### *4.2 Limitations of the study*

This study has several limitations. First, it relies on data collected via the Altmetric, which tracks only Bluesky posts containing direct links to scholarly articles. Posts that mention or discuss articles without including a link are therefore excluded, leading to a degree of data omission. Future research could complement this dataset by employing title-based searches through the Bluesky API, enabling the identification of research-related posts that lack direct links but still engage with scholarly content.

Second, our disciplinary analysis focused on absolute mention counts rather than mention rates, which could be skewed by differences in the volume of publications across fields. A more accurate assessment of disciplinary engagement would involve normalising by the total number of published articles per field and calculating per-article attention metrics.

Finally, we did not analyse the demographic characteristics of Bluesky users. As an emerging platform, Bluesky's user base may differ significantly from that of X, particularly during this transitional period marked by the migration of scientists. It is plausible that current patterns of scholarly discourse are shaped by early academic adopters. Future research should investigate user demographics to better understand the nature and dynamics of scientific engagement on Bluesky.

## 5. Conclusions

This study presents the first large-scale, empirical analysis of scholarly article dissemination on Bluesky, an emerging decentralised platform increasingly adopted by researchers in the post-Twitter era. Drawing on over 2.6 million posts referencing more than 530,000 unique articles, we examined temporal patterns, disciplinary coverage, linguistic and textual characteristics, and user engagement behaviours.

Our findings indicate that Bluesky has evolved into a venue for science communication, driven by a substantial surge in scholarly activity between November 2024 and January 2025, after which engagement levels stabilized at a high baseline. Similar to earlier observations on X, scholarly discourse on Bluesky is characterised by rapid post-publication sharing, disciplinary biases favouring the health, social, and environmental sciences, and the predominance of English. However, unlike Twitter/X, Bluesky posts show a significantly higher degree of interactive engagement and content originality – the former suggesting that scholarly discourse on Bluesky is more participatory and dialogic in nature, and the latter indicating a shift towards more interpretive and reflective forms of scholarly engagement.

These results highlight Bluesky's potential not only as a complementary source for altmetrics but also as a platform that supports meaningful scholarly dialogue. As social media ecosystems continue to evolve, sustained attention to platforms like Bluesky will be essential for understanding the changing nature of public engagement with science and for developing more nuanced, inclusive metrics of research impact.

## Acknowledgements

Zhichao Fang is financially supported by the National Natural Science Foundation of China (No. 72304274). Er-Te Zheng is financially supported by the GTA scholarship from the School of Information, Journalism and Communication of the University of Sheffield. Mike Thelwall is supported by the Fundação Calouste Gulbenkian European Media and Information Fund (No. 316177). The authors thank Altmetric, Bluesky, and OpenAlex for providing access to their APIs, which enabled data collection for this research.

## References

Alperin, J. P., Portenoy, J., Demes, K., Larivière, V., & Haustein, S. (2024). *An analysis of the suitability of OpenAlex for bibliometric analyses* (No. arXiv:2404.17663). arXiv. https://doi.org/10.48550/arXiv.2404.17663

Altmetric team. (2025, January 21). *Bluesky tracking enhancement: An increase to our volume of historical*. https://updates.altmetric.com/announcements/bluesky-tracking-enhancement-an-increase-to-our-volume-of-historical-posts

Arroyo-Machado, W., Robinson-Garcia, N., & Torres-Salinas, D. (2025). Are there stars in bluesky? A comparative exploratory analysis of altmetric mentions between X and bluesky. *Journal of Informetrics*, *19*(3), 101700. https://doi.org/10.1016/j.joi.2025.101700

Bisbee, J., & Munger, K. (2025). The vibes are off: Did elon musk push academics off twitter? *PS: Political Science & Politics*, *58*(1), 139–146. https://doi.org/10.1017/S1049096524000416

Bornmann, L. (2015). Alternative metrics in scientometrics: A meta-analysis of research into three altmetrics. *Scientometrics*, *103*(3), 1123–1144. https://doi.org/10.1007/s11192-015-1565-y

Brossard, D., & Scheufele, D. A. (2013). Science, New Media, and the Public. *Science*, *339*(6115), 40–41. https://doi.org/10.1126/science.1232329

Brown, J., Onik, A. R., & Baggili, I. (2024). Blue skies from (X?s) pain: A digital forensic analysis of threads and bluesky. *Proceedings of the 19th International Conference on Availability, Reliability and Security*, 1–12. https://doi.org/10.1145/3664476.3670904

Campbell, I. C. (2021, August 16). *Twitter's decentralized social network project finally has a leader*. The Verge. https://www.theverge.com/2021/8/16/22627435/twitter-bluesky-lead-jay-graber-decentralized-social-web

Caulfield, M. (2023). The new twitter is changing rapidly—Study it before it's too late. *Nature*, *623*(7986), 225–225. https://doi.org/10.1038/d41586-023-03483-8

Cava, L. L., Aiello, L. M., & Tagarelli, A. (2023). Drivers of social influence in the twitter migration to mastodon. *Scientific Reports*, *13*(1), 21626. https://doi.org/10.1038/s41598-023-48200-7

Coletti, A., McGloin, R., Oeldorf-Hirsch, A., & Hamlin, E. (2022). Science communication on social media: Examining cross-platform behavioral engagement. *The Journal of Social Media in Society*, *11*(2), Article 2.

Costas, R. (2017). Towards the social media studies of science: Social media metrics, present and future. *Bibliotecas. Anales de Investigación*, *13*(1), 1–5.

Costas, R., Zahedi, Z., & Wouters, P. (2015). Do "altmetrics" correlate with citations? Extensive comparison of altmetric indicators with citations from a multidisciplinary perspective. *Journal of the Association for Information Science and Technology*, *66*(10), 2003–2019. https://doi.org/10.1002/asi.23309

Culbert, J. H., Hobert, A., Jahn, N., Haupka, N., Schmidt, M., Donner, P., & Mayr, P. (2025). Reference coverage analysis of OpenAlex compared to Web of Science and Scopus. *Scientometrics*, *130*(4), 2475–2492. https://doi.org/10.1007/s11192-025-05293-3

Didegah, F., Mejlgaard, N., & Sørensen, M. P. (2018). Investigating the quality of interactions and public engagement around scientific papers on Twitter. *Journal of Informetrics*, *12*(3), 960–971. https://www.sciencedirect.com/science/article/pii/S1751157717302572

Erdt, M., Nagarajan, A., Sin, S.-C. J., & Theng, Y.-L. (2016). Altmetrics: An analysis of the state-of-the-art in measuring research impact on social media. *Scientometrics*, *109*(2), 1117–1166. https://doi.org/10.1007/s11192-016-2077-0

Eysenbach, G. (2011). Can Tweets Predict Citations? Metrics of Social Impact Based on Twitter and Correlation with Traditional Metrics of Scientific Impact. *Journal of Medical Internet Research*, *13*(4), e2012. https://doi.org/10.2196/jmir.2012

Fang, Z., Costas, R., Tian, W., Wang, X., & Wouters, P. (2020). An extensive analysis of the presence of altmetric data for Web of Science publications across subject fields and research topics. *Scientometrics*, *124*(3), 2519–2549. https://doi.org/10.1007/s11192-020-03564-9

Fang, Z., Costas, R., & Wouters, P. (2022). User engagement with scholarly tweets of scientific papers: A large-scale and cross-disciplinary analysis. *Scientometrics*, *127*(8), 4523–4546. https://doi.org/10.1007/s11192-022-04468-6

Hassan, S.-U., Saleem, A., Soroya, S. H., Safder, I., Iqbal, S., Jamil, S., Bukhari, F., Aljohani, N. R., & Nawaz, R. (2021). Sentiment analysis of tweets through altmetrics: A machine learning approach. *Journal of Information Science*, *47*(6), 712–726. https://doi.org/10.1177/0165551520930917

Haupka, N., Culbert, J. H., Schniedermann, A., Jahn, N., & Mayr, P. (2024). *Analysis of the Publication and Document Types in OpenAlex, Web of Science, Scopus, Pubmed and Semantic Scholar* (No. arXiv:2406.15154). arXiv. https://doi.org/10.48550/arXiv.2406.15154

Haustein, S. (2016). Grand challenges in altmetrics: Heterogeneity, data quality and dependencies. *Scientometrics*, *108*(1), 413–423. https://doi.org/10.1007/s11192-016-1910-9

Haustein, S., Bowman, T. D., & Costas, R. (2016). Interpreting 'altmetrics': Viewing acts on social media through the lens of citation and social theories. In C. R. Sugimoto (Ed.), *Theories of informetrics and scholarly communication* (pp. 372–405). De Gruyter.

Haustein, S., Costas, R., & Larivière, V. (2015). Characterizing Social Media Metrics of Scholarly Papers: The Effect of Document Properties and Collaboration Patterns. *PLOS ONE*, *10*(3), e0120495. https://doi.org/10.1371/journal.pone.0120495

Haustein, S., Peters, I., Sugimoto, C. R., Thelwall, M., & Larivière, V. (2014). Tweeting biomedicine: An analysis of tweets and citations in the biomedical literature. *Journal of the Association for Information Science and Technology*, *65*(4), 656–669. https://doi.org/10.1002/asi.23101

Hayon, S., Tripathi, H., Stormont, I. M., Dunne, M. M., Naslund, M. J., & Siddiqui, M. M. (2019). Twitter Mentions and Academic Citations in the Urologic Literature. *Urology*, *123*, 28–33. https://doi.org/10.1016/j.urology.2018.08.041

Htoo, T., & Na, J.-C. (2017). Who are tweeting research articles and why? *Journal of Information Science Theory and Practice*, *5*(3), 48–60. https://doi.org/10.1633/JISTaP.2017.5.3.4

Kidambi, M. (2024a, December 3). Altmetric introduces Bluesky as a new social media tracking source. *Altmetric*. https://www.altmetric.com/altmetric-news/altmetric-introduces-bluesky-as-a-new-social-media-tracking-source/

Kidambi, M. (2024b, December 5). Tapping into data from bluesky. *Altmetric*. https://www.altmetric.com/blog/tapping-into-data-from-bluesky/

Kumar, M. S., Gupta, S., Baskaran, S., & Na, J.-C. (2019). User motivation classification and comparison of tweets mentioning research articles in the fields of medicine, chemistry and environmental science. In A. Jatowt, A. Maeda, & S. Y. Syn (Eds.), *Digital Libraries at the Crossroads of Digital Information for the Future* (pp. 40–53). Springer International Publishing. https://doi.org/10.1007/978-3-030-34058-2_5

Kupferschmidt, K. (2024). Researchers and scientific institutions flock to Bluesky. *Science*, *386*(6725), 950–951. https://doi.org/10.1126/science.adu8276

Kupferschmidt, K. (2025). 'Good boring': How Bluesky is shaping scientists' discourse. *Science*, *387*(6730), 123–124. https://doi.org/10.1126/science.adv7962

Mallapaty, S. (2024). 'A place of joy': Why scientists are joining the rush to bluesky. *Nature*, *636*(8041), 15–16. https://doi.org/10.1038/d41586-024-03784-6

McIntyre, K. E. (2014). The evolution of social media from 1969 to 2013: A change in competition and a trend toward complementary, niche sites. *The Journal of Social Media in Society*, *3*(2), Article 2.

McKee, M., Pagel, C., & Buse, K. (2024). Disinformation enabled donald trump's second term and is a crisis for democracies everywhere. *BMJ*, *387*, q2485. https://doi.org/10.1136/bmj.q2485

Na, J.-C. (2015). User motivations for tweeting research articles: A content analysis approach. In R. B. Allen, J. Hunter, & M. L. Zeng (Eds.), *Digital Libraries: Providing Quality Information* (pp. 197–208). Springer International Publishing. https://doi.org/10.1007/978-3-319-27974-9_20

Ortega, J. L. (2018). The life cycle of altmetric impact: A longitudinal study of six metrics from PlumX. *Journal of Informetrics*, *12*(3), 579–589. https://doi.org/10.1016/j.joi.2018.06.001

Peng, H., Romero, D. M., & Horvát, E.-Á. (2022). Dynamics of cross-platform attention to retracted papers. *Proceedings of the National Academy of Sciences*, *119*(25), e2119086119. https://doi.org/10.1073/pnas.2119086119

Peoples, B. K., Midway, S. R., Sackett, D., Lynch, A., & Cooney, P. B. (2016). Twitter Predicts Citation Rates of Ecological Research. *PLOS ONE*, *11*(11), e0166570. https://doi.org/10.1371/journal.pone.0166570

Peters, H. P., Dunwoody, S., Allgaier, J., Lo, Y.-Y., & Brossard, D. (2014). Public communication of science 2.0: Is the communication of science via the “new media” online a genuine transformation or old wine in new bottles? *EMBO Reports*, *15*(7), 749–753. https://doi.org/10.15252/embr.201438979

Priem, J., Groth, P., & Taraborelli, D. (2012). The altmetrics collection. *PLOS ONE*, *7*(11), e48753. https://doi.org/10.1371/journal.pone.0048753

Quelle, D., Denker, F., Garg, P., & Bovet, A. (2025). *Why Academics Are Leaving Twitter for Bluesky* (No. arXiv:2505.24801). arXiv. https://doi.org/10.48550/arXiv.2505.24801

Robinson-García, N., Torres-Salinas, D., Zahedi, Z., & Costas, R. (2014). New data, new possibilities: Exploring the insides of Altmetric.com. *Profesional de la información*, *23*(4), Article 4.

Sergiadis, A. (2018). Analysis of tweets mentioning works from an institutional repository. *Journal of New Librarianship*, *3*(1), 130–150. https://doi.org/10.21173/newlibs/4/27

Sethi, J., Stauss, M., Mirioglu, S., Floyd, L., & Woywodt, A. (2025). Only the bluesky is the limit: Ten tips for a trending #skytorial. *Clinical Kidney Journal*, *18*(2), sfae414. https://doi.org/10.1093/ckj/sfae414

Shuai, X., Pepe, A., & Bollen, J. (2012). How the Scientific Community Reacts to Newly Submitted Preprints: Article Downloads, Twitter Mentions, and Citations. *PLOS ONE*, *7*(11), e47523. https://doi.org/10.1371/journal.pone.0047523

Sugimoto, C. R., Work, S., Larivière, V., & Haustein, S. (2017). Scholarly use of social media and altmetrics: A review of the literature. *Journal of the Association for Information Science and Technology*, *68*(9), 2037–2062. https://doi.org/10.1002/asi.23833

Taylor, M., & Areia, C. (2025). Bluesky's ahead, but is X a dead parrot? *Altmetric*. https://www.altmetric.com/blog/blueskys-ahead-but-is-x-a-dead-parrot/

Thelwall, M., Haustein, S., Larivière, V., & Sugimoto, C. R. (2013). Do altmetrics work? Twitter and ten other social web services. *PLOS ONE*, *8*(5), e64841. https://doi.org/10.1371/journal.pone.0064841

Thelwall, M., Tsou, A., Weingart, S., Holmberg, K., & Haustein, S. (2013). Tweeting links to academic articles. *Cybermetrics*, *17*(1), 1–8.

Traag, V. A., Waltman, L., & Van Eck, N. J. (2019). From Louvain to Leiden: Guaranteeing well-connected communities. *Scientific Reports*, *9*(1), 5233. https://doi.org/10.1038/s41598-019-41695-z

Vaghjiani, N. G., Lal, V., Vahidi, N., Ebadi, A., Carli, M., Sima, A., & Coelho, D. H. (2024). Social media and academic impact: Do early tweets correlate with future citations? *Ear, Nose & Throat Journal*, *103*(2), 75–80. https://doi.org/10.1177/01455613211042113

Vidal Valero, M. (2023). Thousands of scientists are cutting back on twitter, seeding angst and uncertainty. *Nature*, *620*(7974), 482–484. https://doi.org/10.1038/d41586-023-02554-0

Yu, H., Xiao, T., Xu, S., & Wang, Y. (2019). Who posts scientific tweets? An investigation into the productivity, locations, and identities of scientific tweeters. *Journal of*

*Informetrics*, *13*(3), 841–855. https://doi.org/10.1016/j.joi.2019.08.001

Yu, H., Xu, S., & Xiao, T. (2018). Is there Lingua Franca in informal scientific communication? Evidence from language distribution of scientific tweets. *Journal of Informetrics*, *12*(3), 605–617. https://doi.org/10.1016/j.joi.2018.06.003

Yu, H., Xu, S., Xiao, T., Hemminger, B. M., & Yang, S. (2017). Global science discussed in local altmetrics: Weibo and its comparison with Twitter. *Journal of Informetrics*, *11*(2), 466–482. https://www.sciencedirect.com/science/article/pii/S1751157716301961

Zahedi, Z., & Costas, R. (2018). General discussion of data quality challenges in social media metrics: Extensive comparison of four major altmetric data aggregators. *PLOS ONE*, *13*(5), e0197326. https://doi.org/10.1371/journal.pone.0197326

Zahedi, Z., Costas, R., & Wouters, P. (2014). How well developed are altmetrics? A cross-disciplinary analysis of the presence of 'alternative metrics' in scientific publications. *Scientometrics*, *101*(2), 1491–1513. https://doi.org/10.1007/s11192-014-1264-0

Zhang, L., Gou, Z., Fang, Z., Sivertsen, G., & Huang, Y. (2023). Who tweets scientific publications? A large-scale study of tweeting audiences in all areas of research. *Journal of the Association for Information Science and Technology*, *74*(13), 1485–1497. https://doi.org/10.1002/asi.24830

Zhang, Y., & Fang, Z. (2025). The Tower of Babel in science communication on social media: An analysis of linguistic diversity in twitter mentions of scientific publications. *Journal of the Association for Information Science and Technology*. https://doi.org/10.1002/asi.70002

## Appendix

To assess whether the sharp increase in Bluesky scholarly posts from November 2024 to January 2025 reflected a genuine rise in scholarly activity on the platform or was influenced by the commencement of Altmetric's data collection, we employed an alternative data-gathering approach to corroborate the patterns shown in Figure 2. Using the Bluesky API, we retrieved all posts containing links with the string "doi.org", parsed the DOIs embedded in these links, and retained only those whose DOIs could be matched to records in the OpenAlex database. The resulting dataset comprised 126,535 posts authored by 26,306 unique users and citing 102,740 distinct articles.

Although searching posts via the Bluesky API yielded fewer records than collection via Altmetric – likely because posts containing alternative URLs (e.g., publisher pages, institutional repositories, or open-access mirrors) were not captured – it is not constrained by Altmetric's temporal coverage. As posts containing DOIs represent only a subset of all posts citing scholarly articles, temporal changes in this subset can nevertheless serve as an independent indicator of trends in the wider population of Bluesky scholarly posts.

As shown in Figure A1, the numbers of posts, users, and articles exhibit a marked increase in November 2024, with counts of posts and articles rising sharply again in January 2025, followed by relatively stable fluctuations thereafter. The overall trajectory closely parallels that shown in Figure 2, supporting our conclusion that the substantial growth in Bluesky scholarly posts between November 2024 and January 2025 was driven primarily by the large-scale migration of academic users to the platform and their active participation in scholarly discourse, rather than merely by the onset of Altmetric's data collection during this period.

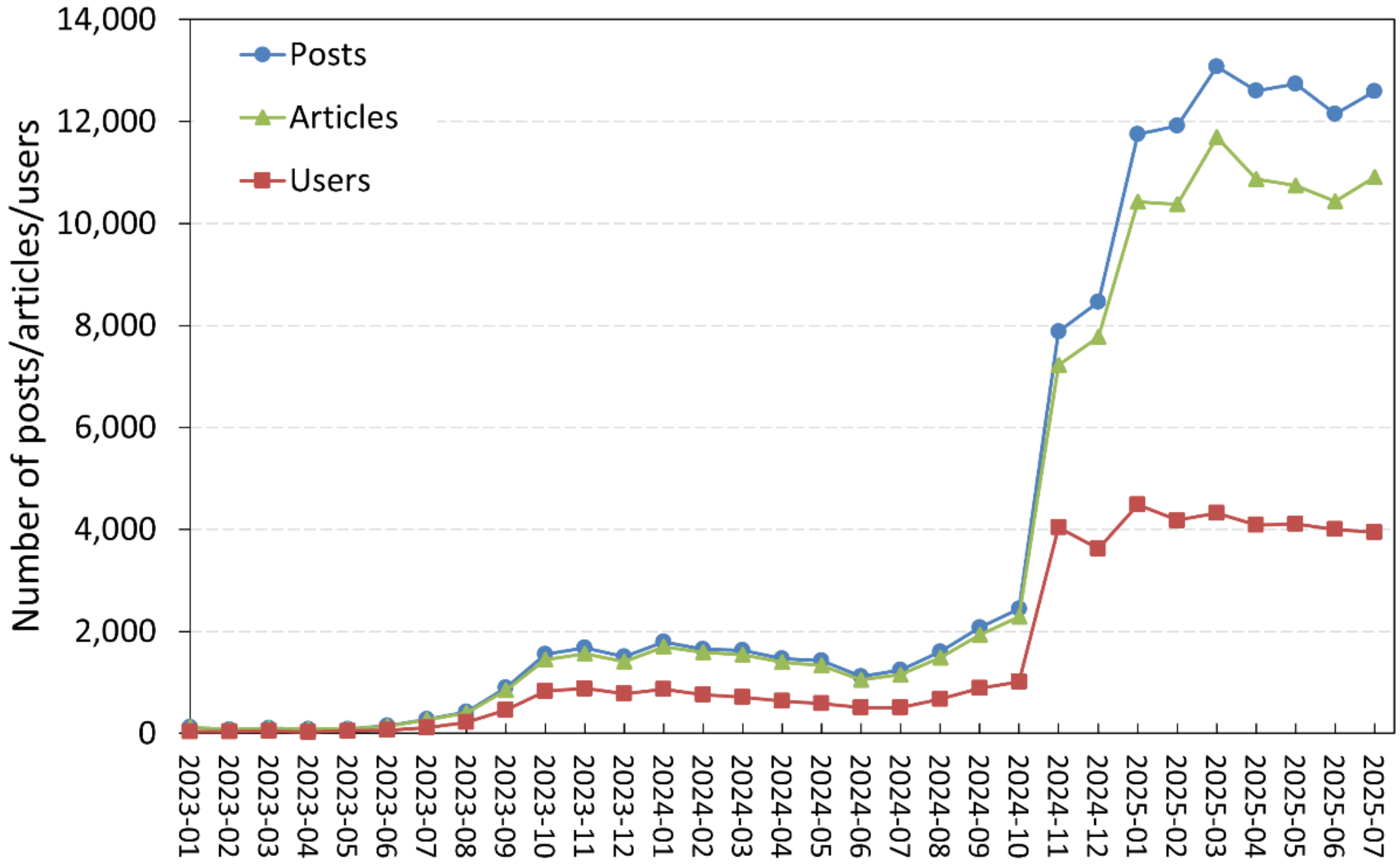

**Figure A1**. Number of Bluesky posts containing valid DOI links, users, and articles referenced from January 2023 to July 2025.